\documentclass[11pt,dvips]{article}

\usepackage{epsfig,times} 
\usepackage{picinpar}
\usepackage{wrapfig}
\usepackage{floatflt}
\makeatletter
\renewcommand{\section}{\@startsection{section}{1}{0in}
	{0.4\baselineskip}{0.1\baselineskip}{\Large\bf}}
\renewcommand{\subsection}{\@startsection{subsection}{2}{0in}
	{0.25\baselineskip}{-\baselineskip}{\large\bf}}
\renewcommand{\subsubsection}{\@startsection{subsubsection}{3}{0in}
	{0.1\baselineskip}{-\baselineskip}{\normalsize\bf}}
\makeatother
\newcommand{\gr}{$\gamma$-ray\ }
\newcommand{\grg}{$\gamma$-ray gradient\ }
\begin{document}

%
\makeatletter\newcommand{\ps@icrc}{
\renewcommand{\@oddhead}{\slshape{}\hfil}}
\makeatother\thispagestyle{icrc}
%
\markright{OG.3.2.19}
\begin{center}
%
{\LARGE \bf The Diffuse Galactic Gamma-Ray Gradient}
\end{center}

\begin{center}
%
%
{\bf D. Breitschwerdt$^{1,2}$, V. Dogiel$^{3}$, and H.J. V\"olk$^{4}$}\\
{\it $^{1}$Heisenberg Fellow\\
$^{2}$Max-Planck-Institut f\"ur extraterrestrische Physik, Postfach 1603,
D-85740 Garching, FRG\\
$^{3}$P.N. Lebedev Physical Institute,  117924 Moscow, Russia\\
$^{4}$Max-Planck-Institut f\"ur Kernphysik, Postfach 103 980, 
D-69029 Heidelberg, FRG}
\end{center}

\begin{center}
{\large \bf Abstract\\}
\end{center}
\vspace{-0.5ex}
%
%
One of the unsolved problems in cosmic ray (CR) physics is the 
small radial gradient of the \gr intensity compared to 
the inferred CR source distribution in the Galactic disk. 
In diffusive CR propagation models the most natural explanation 
is very efficient spatial mixing due to MHD turbulence in the 
interstellar medium. However, even in the most favorable case of 
a very large diffusive CR halo the \grg is still 
larger than observed. In our view the small \grg 
could be the result of strong advection by a Galactic wind. 
We show that a small \grg can be obtained, if the 
diffusion region does {\bf not} extend far above the Galactic plane. 
Important ingredients of our model are: 
(i) anisotropic CR diffusion, 
(ii) strong radial and vertical gradients of the advection velocity 
(due to faster winds above higher CR source density regions). 
%

\vspace{1ex}

%
%
\section{Introduction:}
\label{intro.sec}
The distribution  of the $\gamma$-ray emissivity in the galactic disk bears
important information on the CR origin, because 
Galactic diffuse $\gamma$-rays result from interactions of CRs
with interstellar gas. Thus, this distribution depends critically on the 
conditions of CR propagation in the Galaxy.
For $\gamma$-ray energies above 100 MeV, the main production process is
$\pi^0$-decay, resulting from nuclear collisions between high energy particles
and interstellar matter.
The study of
diffuse $\gamma$-ray emission has shown that the nucleonic component of cosmic
rays is more or less homogeneously distributed over the entire Galactic
disk.
If the spatial distribution of CRs were uniform,
the $\gamma$-ray emission should map perfectly well the distribution of
interstellar hydrogen. However, the distribution of the cosmic ray (CR)
sources (most likely supernova remnants (SNRs) for particle energies below 
$10^{15} \, {\rm eV}$) is far from homogeneous.
Taken at face-value, the discrepancy must arise during the propagation of CRs  
from their sources through the interstellar medium.
The difficulty in interpreting the data correctly is due to the fact, that
the coefficients of CR transport are not well known, and
therefore models (plus their respective assumptions) are used to bridge the gap between
the measured $\gamma$-ray parameters and the unknown parameters of CR origin.

The most common models for the interpretation of 
the spatial CR distribution in the disk are phenomenological in nature and 
based on CR diffusion. It is assumed that particle
propagation in the interstellar medium can be described by an isotropic diffusion
process due to CR scattering off magnetic field fluctuations. The value of the 
diffusion coefficient (or tensor) is estimateded from observational data. 
In many cases this model gives an acceptable interpretation of experimental 
results (see e.g.\ Berezinsky et al., 1990).

The interpretation of the \grg in the framework of an
isotropic diffusion model, in which the CRs are injected by a radially non-uniform 
SNR distribution (e.g.\ Case \& Bhattacharya, 1996) leads 
to the following conclusion: in order to obtain a rather uniform CR
distribution in the disk one has to assume that the Galaxy is surrounded by a
huge halo whose vertical extension is larger than 10 kpc (Bloemen et al., 1993).
Only then diffusion mixes CRs efficiently enough, so that their distribution in the
disk differs appreciably from that of their sources. The numerical calculations 
(see Dogiel \& Uryson, 1988) show however, that even in the case of a {\it very
extended diffusion halo} the calculated gradient of CRs is larger than derived
from COS-B and EGRET data. Therefore, if we believe that the observations of
the SNR and $\gamma$-ray distributions are correct, we should seek alternative 
explanations for the gradient data.

\section{Effects of CR Transport by a Galactic Wind:}
\label{crtran.sec}
A completely different scenario of CR escape into intergalactic space
is implied by the so-called galactic wind models. It is shown that the
combined pressure of thermal plasma, CRs, magnetic fields and MHD waves
can lead to a secular escape of gas and CRs in normal spiral galaxies,
provided that the coupling between scattering waves and energetic nucleons
is strong. In this case there is a net forward momentum transfer from CRs to the
gas via the frozen-in waves as a mediator (see  Breitschwerdt et al., 1991;
1993).
In wind models the coupling between MHD-waves and CRs is generally assumed to be 
strong everywhere in space except for a rather narrow region surrounding 
the galactic plane where CRs propagate predominantly by diffusion. 
In the wind CRs are transported along with the gas and the waves 
to large distances from the disk. As numerical simulations have shown, 
these models exhibit 
two effects that may provide the long-sought explanation for the $\gamma$-ray
gradient variation problem, both of which are not present in static halo models.
The first one is that the diffusion tensor may be
{\it anisotropic}. The reason is that the component along the magnetic field lines
depends on the CR energy, as $D_\parallel \propto E^{0.6}$, and the component
perpendicular to the lines $D_\perp$ is energy independent. As a result in
the case of a strong galactic wind the location, $z_{\rm c}$, of the halo boundary
depends on the energy as: $z_{\rm c} \propto E^{0.6}$
and the local CR life time $T_{cr}\sim {{z_{\rm c}^2}\over D}\propto E^{0.6}$.
The diffusion component $D_\parallel$ rises with energy so the diffusion is one
dimensional at high energies.

The second, and more important, new effect is the {\it radial variation of the 
wind velocity} (or mass flux, if the gas density is constant) of the galactic wind 
along the galactic plane, which may give rise to an 
almost uniform CR distribution in the disk, even for a strongly non-uniform 
CR source distribution, provided that the diffusion halo height is small. 
The reason is that, depending on the local CR source strength (i.e.\ the
number of SNRs per unit volume and hence the number density of CRs at a given
energy, $n(r, z, E)$), the CR energy density, $\epsilon_{\rm c}$ or its
pressure $P_{\rm C}$, will vary
accordingly with galactocentric radius. Thus an increase in $\epsilon_{\rm c}$
will also increase the galactic wind mass flux or wind velocity, respectively,
and hence decrease the distance to the advection boundary, given by 
${D({R_{\mathrm{c}}}, {z_{\mathrm{c}}}, E)}/
({V({R_{\mathrm{c}}}, {z_{\mathrm{c}}}){z_c}}) \sim 1$, where $R$ and 
$V=u+V_{\rm A}$ 
denote the halo radius and the CR transport speed with respect to an Eulerian
frame of reference. 
Consequently, the CR storage volume will be reduced,
thus facilitating CR escape locally and thereby decreasing the number of nuclear
collisions, which produce $\gamma$-rays via $\pi^0$-decay.

\section{Uniform Disk Distribution of CRs in the Wind Model:}
\label{crdis.sec}
To illustrate the behaviour of the solution of the advection-diffusion
equation we discuss two simple cases. The first one describes one-dimensional
diffusion with the advection velocity $V$ depending on the radial coordinate 
$r$ only, as $V(r)= V_0\cdot f_1(r)$, 
with $V_0$ being a constant and $f_1(r)$ an arbitrary function of $r$.
The equation to solve for the CR distribution function $N(r,z)$ is
\begin{equation}
-{d\over {dz}}\left( D{{dN}\over {dz}} -V(r) N\right)=Q \cdot f_2(r)
\label{diff-eq1}
\end{equation}
with the boundary condition $N=0$ at $z_c=1$. Here $Q$ is the non-radial part of the 
CR source distribution, and 
$f_2(r)$ is also an arbitrary function of $r$, which describes the radial 
SNR distribution in the disk. The solution of Eq.~(\ref{diff-eq1}) is
\begin{equation}
N(z,r)={{Q f_2(r)}\over{V_0 f_1(r)}}\left(1 - e^{-(V_0 f_1/D)(1-z)}\right) \,.
\end{equation}
We see that for the case of weak advection $V_0 \ll 1$
the solution is $N\approx Q f_2 (1-z)/D$, 
and the CR gradient depends on the source distribution $f_2$ only.
However if $V_0 \gg 1$ then
\begin{equation}
N \approx {{Q f_2(r)}\over {V_0 f_1(r)}} \,,
\end{equation}
and in this case the gradient is a strong function of the velocity distribution too. 
In the special case $f_1(r)=f_2(r)$ we obtain $N=const$.

Let us investigate a more complex case in which the velocity is also a function 
of $z$. For a velocity distribution $V(z,r) =3\, V_0 \,z \, f_1(r)$, with some 
suitable constant $V_0$,  
the solution of the one-dimensional diffusion equation was obtained by Bloemen
et al., 1993. 
For strong advection the solution converges to
\begin{equation}
N(z=0,r)\simeq A{{Q f_2(r)}\over{\sqrt{D V_0 f_1(r)}}} \,;
\end{equation}
we retrieve $N=const$, if $f_1(r) =f_2^2(r)$, where $A$ is a constant.

The diffusion coefficient $D$ is assumed to be constant, although we cannot
exclude its spatial variation, which can change the CR density distribution in
the disk as well.

\section{Analytical Solution of the Diffusion-Advection Equation:}
\label{ansol.sec}
We now want to work out a general solution of the two-dimensional CR transport
equation for nucleons, which reads 
%
\begin{equation}
- \nabla \left(\underline{\underline D}(\vec{x}, E) \nabla N -
\vec{u}(\vec{x}) N \right) - {\partial \over \partial E} \left(\frac{1}{3}
\nabla\vec{u}(\vec{x}) \, E \, N \right)  
- {\partial \over \partial E} \left({d E \over dt} \, N\right) = Q(E, \vec{x}) \,,
\label{2dcon-diff1}
\end{equation}
where $\vec{x}$ denotes the spatial coordinates, and the diffusion tensor 
$\underline{\underline D}$ in cylindrical coordinates
with axial symmetry is given by
\begin{equation}
\underline{\underline D} = \left(
\begin{array}{cc}
D_{rr} & D_{rz} \\
D_{zr} & D_{zz}
\end{array}
\right)
= \left(
\begin{array}{cc}
\kappa_{r} & 0 \\
0 & D_{z}\cdot E^{\alpha}
\end{array}
\right) \,.
\end{equation}
Our prime interests here are the anisotropic diffusion and advection, 
rather than the energy dependence
and so, for convenience, we set $\alpha =0$.
For the nucleon component other than adiabatic losses are negligible
($dE/dt=0$); thus Eq.~(\ref{2dcon-diff1}) in axial symmetry becomes
\begin{equation}
- D_z {\partial^2 N \over \partial z^2} - \kappa_r
\left({\partial^2 N \over \partial r^2} + \frac{1}{r} {\partial N \over \partial r}
\right) + 3 {V_0 \over r^2} {\partial \left(N \, z\right) \over \partial z} 
- {V_0 \over r^2} {\partial \left(N \, E \right) \over \partial E} =
Q_1(r) \, \delta(z) \, E^{-\gamma_0} \,,
\label{2dcon-diff2}
\end{equation}
where we have chosen a spatial variation for the velocity field as
\begin{equation}
\vec{u}(\vec{x}) = \left(
\begin{array}{c}
u_r \\
u_z
\end{array}
\right)
= \left(
\begin{array}{c}
0 \\
3 {V_0} z/{r^2} 
\end{array}
\right) \,,
\end{equation}
and we have used a power law spectrum for particle injection by the disk
sources. The choice of $\vec{u}(\vec{x})$ seems rather arbitrary at first
glance, and it can indeed be quite a complicated function of $r$ and
$z$. However, numerical calculations of galactic winds (Breitschwerdt et al.,
1991) show a similar spatial variation of the
velocity field like the one we have chosen.

The full analytical solution od Eq.~(\ref{2dcon-diff2}) is rather cumbersome, 
and will be discussed in detail elsewhere (Breitschwerdt et al., 1999). 
Here we present the result for the CR distribution function in the disk ($z=0$), 
subject to the boundary conditions $N(r,z=\pm\infty)= 0$ and 
$N(r=\pm\infty,z)= 0$:
\begin{eqnarray}
&& N(r, z=0, E) = \mp \frac{1}{4 \pi^3 D_z}
E^{-\gamma_0} \Biggl\{\int_{\bar r}^{+\infty} Q_1(r^{\prime}) 2 \pi r^\prime
dr^\prime \sum_{n=0}^{\infty} {{(-1)^n
 \Gamma\left(\frac{1-\beta_n}{2}\right)}
 \over
{n!\Gamma\left(\frac{1-2n}{2}\right)
 \Gamma\left(1-\frac{\beta_n}{2}\right)}}\times
\nonumber \\
&\times&
\left\{(-1)^{\tilde\gamma_n} \cos\left(\frac{\pi\beta_n}{2}\right)
\over
(-1)^{\tilde\gamma_n} \sin\left(\frac{\pi\beta_n}{2}\right)+ (-1)^n
\right\} \,
\left({\bar r}\over r^{\prime}\right)^{-K_n} {{dK_n}\over {d\alpha}}+
\int_{0}^{\bar r} Q_1(r^{\prime}) 2 \pi r^\prime dr^\prime
\sum_{m=0}^{\infty}
{{(-1)^m
 \Gamma\left(\frac{\alpha_m}{2}\right)}
 \over
{m!\Gamma\left(\frac{1-2m}{2}\right)
 \Gamma\left(1+\frac{\alpha_m}{2}\right)}}\times
\nonumber \\
&\times&
\left\{
\sin\left(\frac{\pi\alpha_m}{2}\right)
\over
(-1)^{\tilde\gamma_m+m} \sin\left(\frac{\pi\beta_n}{2}\right) +
\cos\left(\frac{\pi\alpha_m}{2}\right)
\right\}
\left({\bar r}\over r^{\prime}\right)^{-K_m} {{dK_m}\over {d\beta}}\Biggr\}\,.
\label{disk-sol1}
\end{eqnarray}
Here,  
$\beta_n=-2n-3A-2\sqrt{A(\gamma + 6n)}$, 
$\tilde\gamma_n = {1\over 2}(1-4n - 3 A - 2\sqrt{A(\gamma+6n)})$, 
$\tilde\gamma_m = {1\over 2}(2(2m+1)-3 A+1+2\sqrt{A(\gamma-3(2m+1))})$, 
$K_n = - 2n - \sqrt{A(\gamma + 6n)}$, 
${dK_n}/{d\alpha} = 1 + 3 A/(2\sqrt{A(\gamma+6n)})$, 
$\alpha_m = 2m + 1 - 3A + 2 \sqrt{A(\gamma - 3(2m+1))}$, 
$K_m = 2m+1+\sqrt{A(\gamma-3(2m+1))}$, 
${dK_m}/{d\beta} = 1 - 3 A/(2\sqrt{A(\gamma-3(2m+1))})$, where 
$n, m =0,1,2,...$, $D_z =D_\perp$,  
$A=V_0/\kappa_r$, $B=\sqrt{D_z/\kappa_r}$, $\bar r = B r$, and 
$\gamma = \gamma_0+2$. 
 
It is instructive to look at the asymptotic behaviour of the solution at different 
radii. If the CR sources
occupy a limited volume of the disk bounded by a radius $a$, and we let 
$Q(r^\prime )= Q r'^{-q} \, \Theta (a-r^\prime)$ (where $q$ is a fit to the observed 
SNR distribution, and $\Theta$ denotes the step function), then for $r>a$ the 
function $N$ is completely determined by the second integral which is a constant. 
Indeed from contour integration we find that the first
integral ${\int_{r}^{+\infty}}=0$ and the second integral
${\int_{0}^r}={\int_{0}^a}=const$ for $r>a$. Then from a simple analysis we
see that
$N(r)\propto r^{-1-\sqrt{A(\gamma-3)}}$
and
$ N(r)\propto r^{-1}$
if $A\ll 1$. This is just what is expected for purely diffusive particle propagation. 
In the case of strong advection, i.e.\ $A\gg 1$, we have 
$
N(r)\propto r^{-\sqrt{A(\gamma-3)}}
$;
such a strong drop of CR density far
away from the sources is due to adiabatic energy losses, since, if we
neglect them, $\gamma=3$ formally, and $ N(r)\propto r^{-1}$, 
almost independent of the value of $A$.

In the vicinity of the source region ($r\ll a$) the function $N$ is determined
by the two integrals (cf.\ Eq.~(\ref{disk-sol1})) which can be written as
\begin{equation}
N(r)\propto r^{\sqrt{A\gamma}}{\int_r^{+\infty}}{r'}^{(1-q-\sqrt{A\gamma})}\, 
dr^\prime - r^{-1-\sqrt{A(\gamma-3)}}{\int_0^r}{r'}^{(1-q+\sqrt{A(\gamma-3)})}
dr^\prime \,.
\end{equation}
For the case of strong advection
$A\gg 1$ the first integral is determined by the lower limit and the second one
by the upper limit, {\em independent} of the source distribution. Then we have
$
N(r)\propto r^{2-q}
$
and the CR density is almost in the disk region close to the sources if $q=2$.

If $A\ll 1$ then the values of the integrals are determined by the source
distribution. If the CR sources are concentrated towards the centre of the disk, 
$q>2$, then
the CR distribution is determined by the second integral and
$ N(r)\propto r^{-1-\sqrt{A(\gamma-3)}}$. 
If the sources are uniformly distributed in the
disk ($q=0$) then
$ N\propto (a^2 - r^2)$. 
Thus from these analyses we see that a more or less uniform CR distribution 
can be expected in the disk, if advection is strong, i.e.\ $A\gg 1$.

\section{Conclusion:}
\label{conc.sec}
If the advection velocity is proportional
to the CR source density in the disk, then the propagation 
characteristics of the CRs at each point of the disk are strongly determined by local
conditions. In this case we can obtain a radially uniform CR distribution, 
if the halo boundary is close to the disk.  Therefore 
local CR characteristics should  vary strongly from point to point in
the disk.  Moreover, any attempt to estimate global parameters of CRs (such as
the halo height or diffusion coefficient) from  distribution of their density
in the disk is misleading, since mixing of CRs in the Galaxy may be rather weak.
Therefore one may conclude that in the case of strong CR advection, the halo 
extension derived from CR nuclear data reflects only a local halo extension 
near Earth and the value derived from \gr data and pure diffusion models may be an 
artifact.
%
%
%
%
%
%
\vspace{1ex}
\begin{center}
{\Large\bf References}
\end{center}
%
Berezinsky, V.S., Bulanov, S.V., Dogiel, V.A., Ginzburg, V.L., \& Ptuskin, V.S. 
1990, Astrophysics of Cosmic Rays, (ed.\ V.L.\ Ginzburg), North Holland\\
Bloemen, J.B.G.M., Dogiel, V.A., Dorman, V.L., \& Ptuskin, V.S. 1993,
A\&A 267, 372\\
Breitschwerdt, D., Dogiel, V.A., \& V\"olk, H.J. 1999, A\&A (in preparation)\\
Breitschwerdt, D., McKenzie, J.F., \& V\"olk, H.J. 1991, A\&A 245, 79\\
Breitschwerdt, D., McKenzie, J.F., \& V\"olk, H.J. 1993,   A\&A 269, 54\\
Case, G., \& Bhattacharya, D. 1996, A\&AS 120, 437\\ 
Dogiel, V.A., \& Uryson, A.V.\ 1988, A\&A 197, 335\\

\end{document}